\documentclass{IEEEtran}
\usepackage{cite}
\usepackage{amsmath,amssymb,amsfonts}
\usepackage{graphicx}
\usepackage{textcomp,nicefrac}
\def\BibTeX{{\rm B\kern-.05em{\sc i\kern-.025em b}\kern-.08em
T\kern-.1667em\lower.7ex\hbox{E}\kern-.125emX}}
\usepackage[colorlinks,linkcolor=blue,anchorcolor=blue,citecolor=blue]{hyperref}
\usepackage[switch]{lineno}

\usepackage[LGRgreek]{mathastext}
\MTgreekfont{cmr}
\mathastext[upright]

\begin{document}

\title{Impact of Gain Layer Doping Concentration on the Performance of 4H-SiC LGADs}
\author{Tao Yang, Ben J. Sekely, Yashas Satapathy, Abraham Tishelman-Charny, Greg Allion, Gil Atar, Philip Barletta, Gabriele Giacomini, Carl Haber, Steve Holland, John F. Muth, Spyridon Pavlidis, Chengxi Yang, Haichen Wang and Stefania Stucci
\thanks{This work is supported by the U.S. Department of Energy, Office of Science, Office of High Energy Physics, under Contract Number DE-AC02-05CH11231 and Award Number DE-SC0024252. Additional support was provided by the National Science Foundation under contract No. ECCS-1542015.}
\thanks{Tao Yang, Steve Holland and Carl Haber (e-mail: chhaber@lbl.gov) are with Physics Division, Lawrence Berkeley National Laboratory, 1 Cyclotron Road, Berkeley, 94720, CA, U.S.}
\thanks{Ben J. Sekely, Yashas Satapathy, Greg Allion, Gil Atar, Philip Barletta, Spyridon Pavlidis and John F. Muth (e-mail: muth@ncsu.edu) are with Department of Electrical and Computer Engineering, North Carolina State University, 2410 Campus Shore Dr, Raleigh, 27606, NC, U.S. The current affiliation of Ben J. Sekely is NRL/National Research Council, Washington, DC 20375, U.S.}
\thanks{Chengxi Yang and Haichen Wang are with Physics Department, University of California, Berkeley, 94720, CA, U.S.}
\thanks{Abraham Tishelman-Charny, Gabriele Giacomini and Stefania Stucci are with Brookhaven National Laboratory,  Upton, 11973, NY, U.S.}
}

\maketitle


\begin{abstract}
Gain layer doping concentration determines the electric field profile in a low gain avalanche detector (LGAD) and is therefore the primary design parameter. This work reports its influence on the charge collection and timing behavior of 4H-SiC LGADs. Two mesa-isolated devices with cold He ion implanted termination differ only in gain layer doping, $4\times10^{17}$~$cm^{-3}$ and $5\times10^{17}$~$cm^{-3}$, and are characterized together with a PIN diode from the same wafer using the ultraviolet transient current technique (UV-TCT). The 25\% doping difference produces a gain difference of about two orders of magnitude at the same applied bias, representing a gain range of 3.7 to 216. Under UV-TCT excitation, the higher-doped device exhibits time-resolution minima within a gain range of about 20--80 across the investigated laser intensities, consistent with the transition from electronic-noise-dominated jitter to multiplication shot noise. The non-monotonic timing response is also observed under a $^{90}Sr$ beta source with a Si LGAD reference: the higher-doped device reaches 51.7 ps at 280 V and then degrades, while the lower-doped device requires more than 450 V. This measured gain window provides an experimental reference for further optimization of the gain-layer design.

\end{abstract}


\section{Introduction} \label{sec:introduction}

4H-SiC has emerged as an attractive material for particle detectors in future high-energy physics experiments owing to its low leakage current, high carrier saturation drift velocity, and excellent thermal stability \cite{DeNapoli_SiC_Review}. Motivated by the success of silicon-based LGADs \cite{Si_LGAD_2022}, 4H-SiC LGADs employ internal gain to compensate for the small ionization charge deposited by minimum ionizing particles in SiC \cite{Tao_Design_SiC_LGAD}, and several groups have reported working devices \cite{Tao_EDL,SVIHRA_SiC_LGAD_2025,Zhao_SiC_LGAD_2025}. Building on these efforts, this work focuses on how the choice of the gain layer doping itself affects the achievable timing performance.

The gain layer doping concentration is a key parameter governing this performance, since it sets the peak electric field at the multiplication junction and hence the attainable gain \cite{Book_UFSD}. Because the impact ionization rate depends exponentially on the electric field, a small change in doping produces a large change in gain at a given bias, which in turn shifts the operating voltage, the signal-to-noise ratio, and ultimately the timing performance. The most direct way to examine where the time resolution is optimized along the gain axis is therefore to use a set of devices that differ only in doping and jointly cover a sufficiently wide gain range. This work uses exactly such a pair: two 4H-SiC LGADs identical except for the gain layer doping ($5\times10^{17}$ versus $4\times10^{17}$~cm$^{-3}$), characterized in parallel with a PIN diode taken from the same epitaxial wafer as a unity-gain reference using a UV-TCT system. Comparing the two devices over a wide bias range and at several injected charge levels allows the time resolution to be traced continuously from unity gain to above 200, and the resulting gain dependence is interpreted with a signal-to-noise model. This comparison aims to establish the gain window that can realistically be targeted in the design of the 4H-SiC gain layer.

\section{Device Structure} \label{sec:device_structure}

\begin{figure}[!t]
\centerline{\includegraphics[width=1.0\columnwidth]{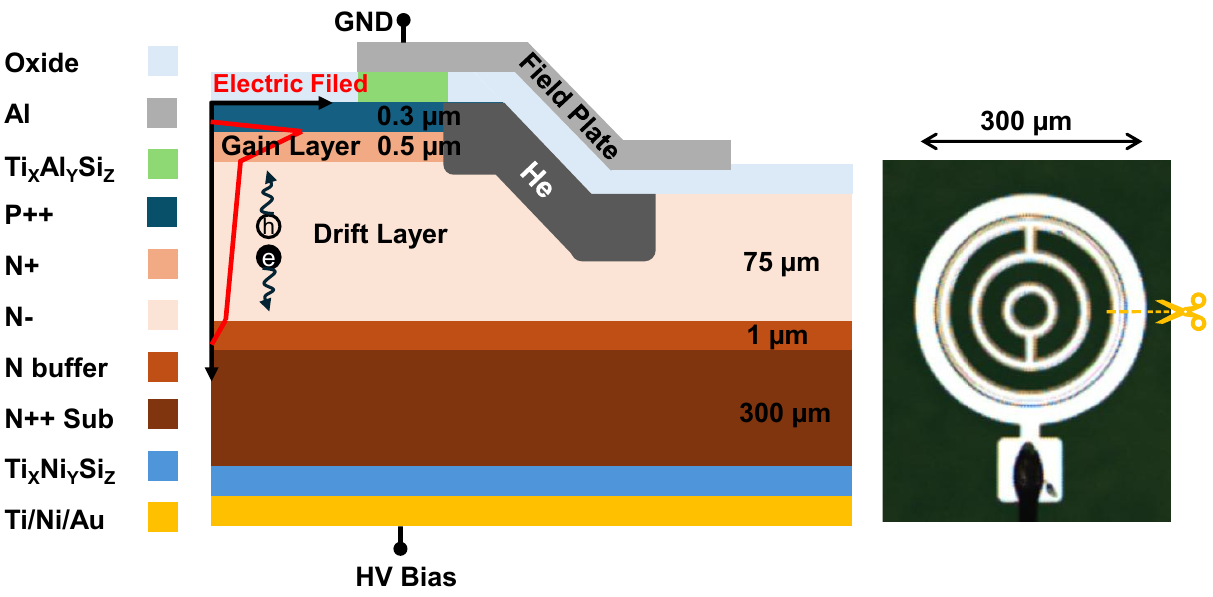}}
\caption{Left: Cross-sectional sketch of the mesa-isolated 4H-SiC LGAD with cold He ion-implanted sidewall termination and field plate. The red line sketches the electric field profile along the mesa center; Right: a micrograph of the device shown as an inset.}
\label{fig:device}
\end{figure}

\begin{figure}[!t]
\centerline{\includegraphics[width=1.0\columnwidth]{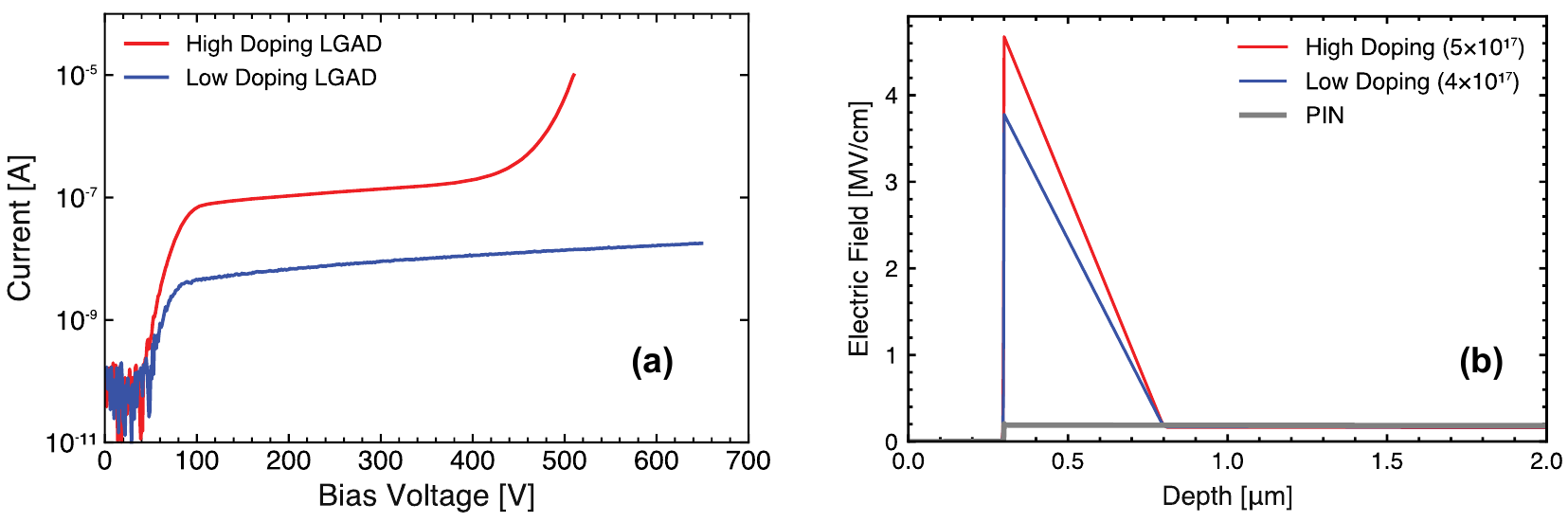}}
\caption{(a)~Reverse bias I-V characteristics of LGADs with both low ($4\times10^{17}$~cm$^{-3}$) and high ($5\times10^{17}$~cm$^{-3}$) gain layer doping levels at room temperature; (b)~Electric field distribution in the two LGADs and PIN diode at 500~V bias.}
\label{fig:iv_ef}
\end{figure}

The devices studied in this work are mesa-isolated 4H-SiC LGADs whose edge termination combines cold He ion implantation along the mesa sidewall \cite{Ben_ID_LGAD_Talk,Sekely2025SiCLGAD} with a metal field plate \cite{Tao_EDL}, as shown on the left side of Fig.~\ref{fig:device}. The acceptor-like defects introduced by the implantation compensate the n-type dopants in the implanted region and thereby mitigate the electric field crowding at the mesa edge without requiring high-temperature post-implantation annealing. Both LGADs have a diameter of 300~$\mu$m and share the same epitaxial structure: a 0.3~$\mu$m p$^{++}$ contact layer ($1\times10^{20}$~cm$^{-3}$), a 0.5~$\mu$m n-type gain layer, and a 75~$\mu$m n$^{-}$ drift layer ($2-4\times10^{14}$~cm$^{-3}$). The only difference between the two devices is the gain layer doping concentration: $5\times10^{17}$~cm$^{-3}$ for the high-doping LGAD and $4\times10^{17}$~cm$^{-3}$ for the low-doping LGAD. A PIN diode fabricated from the same epitaxial stack without the gain layer provides the unity-gain reference. Given the drift-layer doping concentration of $2$--$4\times10^{14}$~cm$^{-3}$, full depletion of the 75~$\mu$m drift layer requires a bias of approximately 1050~V, which is beyond the bias ranges investigated in this work. Both LGADs and the PIN diode therefore remain partially depleted throughout the measurements.

The reverse biased I-V characteristics are shown on the left side of Fig.~\ref{fig:iv_ef}(a). The lower-doped LGAD remains below 20~nA up to 650~V with no sign of breakdown. The higher-doped LGAD exhibits a higher leakage plateau of about 0.1~$\mu$A, begins to show soft breakdown near 450 V, and reaches the 10~$\mu$A compliance at 510 V. Owing to the wide bandgap of 4H-SiC, bulk thermal generation is negligible at room temperature. The measured leakage current is therefore attributed mainly to surface and sidewall leakage along the mesa termination, possibly together with defect-related generation in the epitaxial layer; the part of this current that flows through the gain layer is multiplied, consistent with the higher leakage plateau of the higher-doped LGAD. Due to the differences in the doping of the gain layer, the electric field within the gain layer also differs under the same bias voltage. Fig.~\ref{fig:iv_ef}(b) shows the electric field distribution in the gain layer obtained from simulations under the same bias. An electric field $>$3~MV/cm can trigger carrier multiplication as they drift into the gain layer. The electric field increases with the applied bias voltage, eventually leading to breakdown. The UV-TCT measurements are therefore limited to 375 V for the higher-doped LGAD and 500 V for the lower-doped LGAD. These limits also keep both devices away from the bias range close to breakdown, where the rapidly rising dark count rate would introduce additional noise into the amplitude and timing measurements.


\section{Test Setup} \label{sec:setup}

\begin{figure}[!t]
\centerline{\includegraphics[width=1.0\columnwidth]{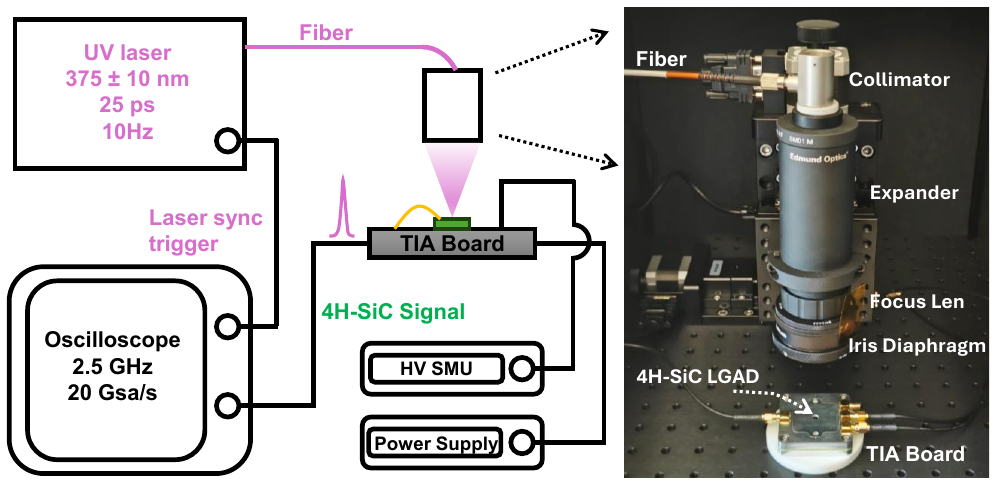}}
\caption{Schematic of the UV-TCT system (left) and the components of the focusing optics (right).}
\label{fig:setup}
\end{figure}

\begin{figure}[!t]
\centerline{\includegraphics[width=1.0\columnwidth]{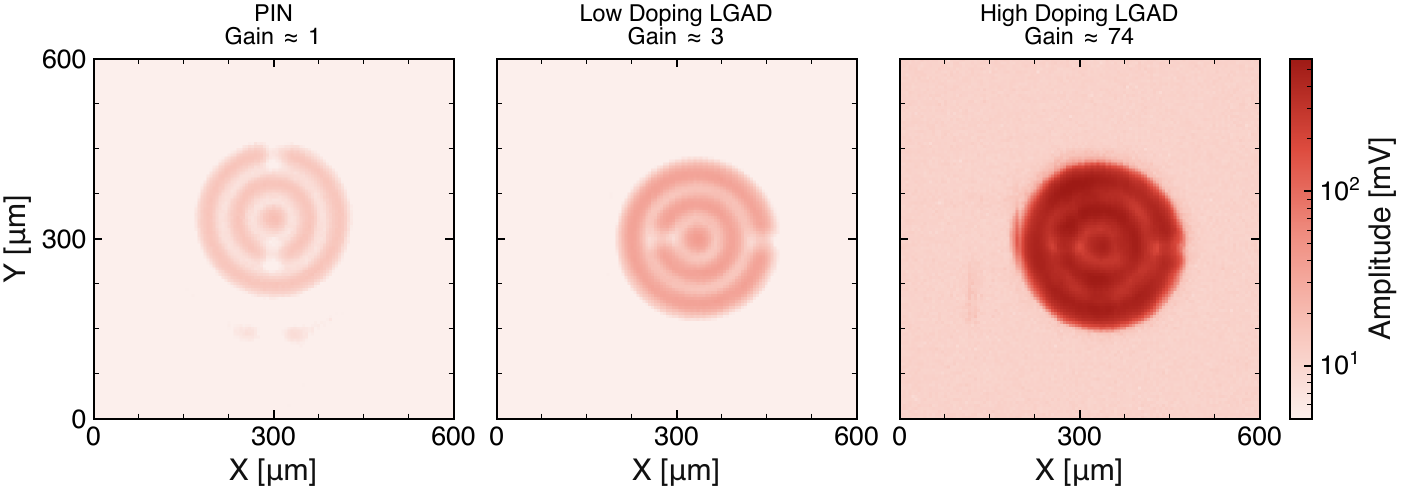}}
\caption{Signal amplitude maps obtained from UV-TCT scans in the $xy$ plane at 300~V. From left to right: PIN, low-doping LGAD, and high-doping LGAD, with the gain at 300~V indicated above each map.}
\label{fig:map}
\end{figure}    

\begin{figure}[!t]
\centerline{\includegraphics[width=0.98\columnwidth]{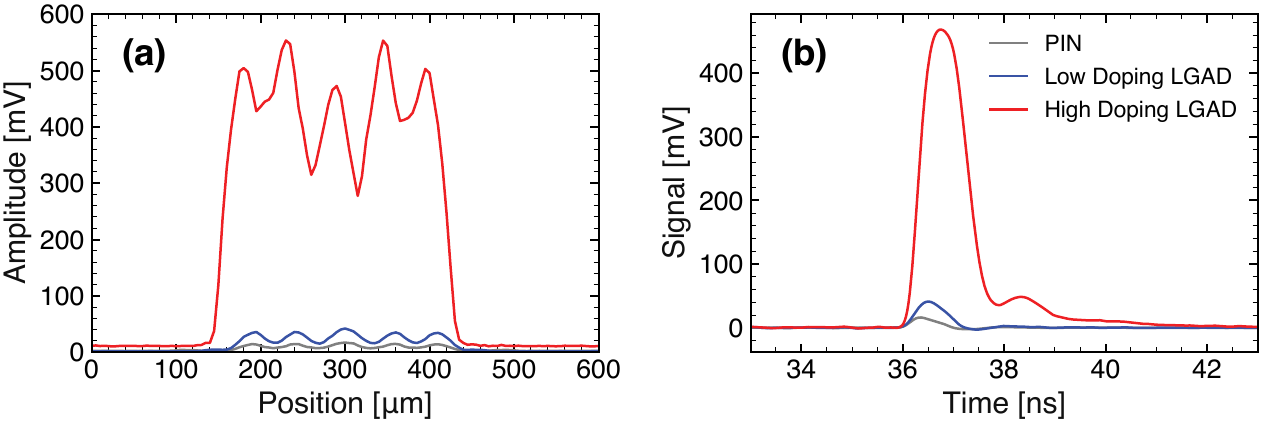}}
\caption{(a) Signal amplitude along a line scan through the device center (extracted from data presented in Fig. \ref{fig:map}); (b) laser-induced pulses of the three devices at a bias voltage of 300~V.}
\label{fig:scan}
\end{figure}

The UV-TCT setup used in this work is identical to that reported in \cite{Tao_EDL}, as shown in Fig.~\ref{fig:setup}. A 375~nm ultraviolet laser with a pulse width of 25~ps is coupled into a multimode optical fiber and delivered to a collimator. The beam is then expanded, focused, and passed through an aperture to form a spot with an effective intensity diameter of approximately 50~$\mu$m, estimated from Gaussian fits to knife-edge measurements. The focused laser pulse impinges on the 4H-SiC LGAD and generates a transient current pulse, which is read out by a high-speed transimpedance amplifier board and recorded by a high-bandwidth oscilloscope (2.5~GHz, 20~GSa/s). The synchronous trigger output of the laser is recorded on the second oscilloscope channel and serves as the time reference for the timing measurements in Sec.~\ref{sec:timing}. Both the repetition rate and the intensity of the laser are adjustable in this work. A repetition rate of 10~Hz is chosen to avoid damage to the 4H-SiC LGAD from prolonged high-repetition-rate illumination and to allow carriers trapped at defects in the material to be released between adjacent pulses, so that the accumulated trapped charge does not perturb the internal electric field from pulse to pulse. The laser intensity is increased stepwise starting from the lowest level at which a pulse signal can just be observed in the high-doping LGAD, and the injected charge is calibrated with the 4H-SiC PIN diode at the intensities where its signal can be resolved. Four intensity settings, denoted I1 to I4 in order of increasing injected charge, are used throughout this work, where the signal differs by about one order of magnitude between I1 and I4.

The z position is first adjusted with a motorized xyz stage of micrometer resolution to obtain the optimum focus. The device is then scanned in the xy plane to obtain the spatial distribution of the signal response shown in \figurename~\ref{fig:map} and \figurename~\ref{fig:scan}. At the same laser intensity and bias, both LGADs with different doping produce signal amplitudes higher than that of the PIN diode, with the higher-doped LGAD exhibiting the largest gain and the corresponding signal amplitude. The images reflect the shadowing effect of the metal electrode on the ultraviolet laser: the detectable signal response is confined to the window region of the device surface not covered by the electrode.

\section{Charge Collection} \label{sec:charges_collection}

\begin{figure}[!t]
\centerline{\includegraphics[width=1.0\columnwidth]{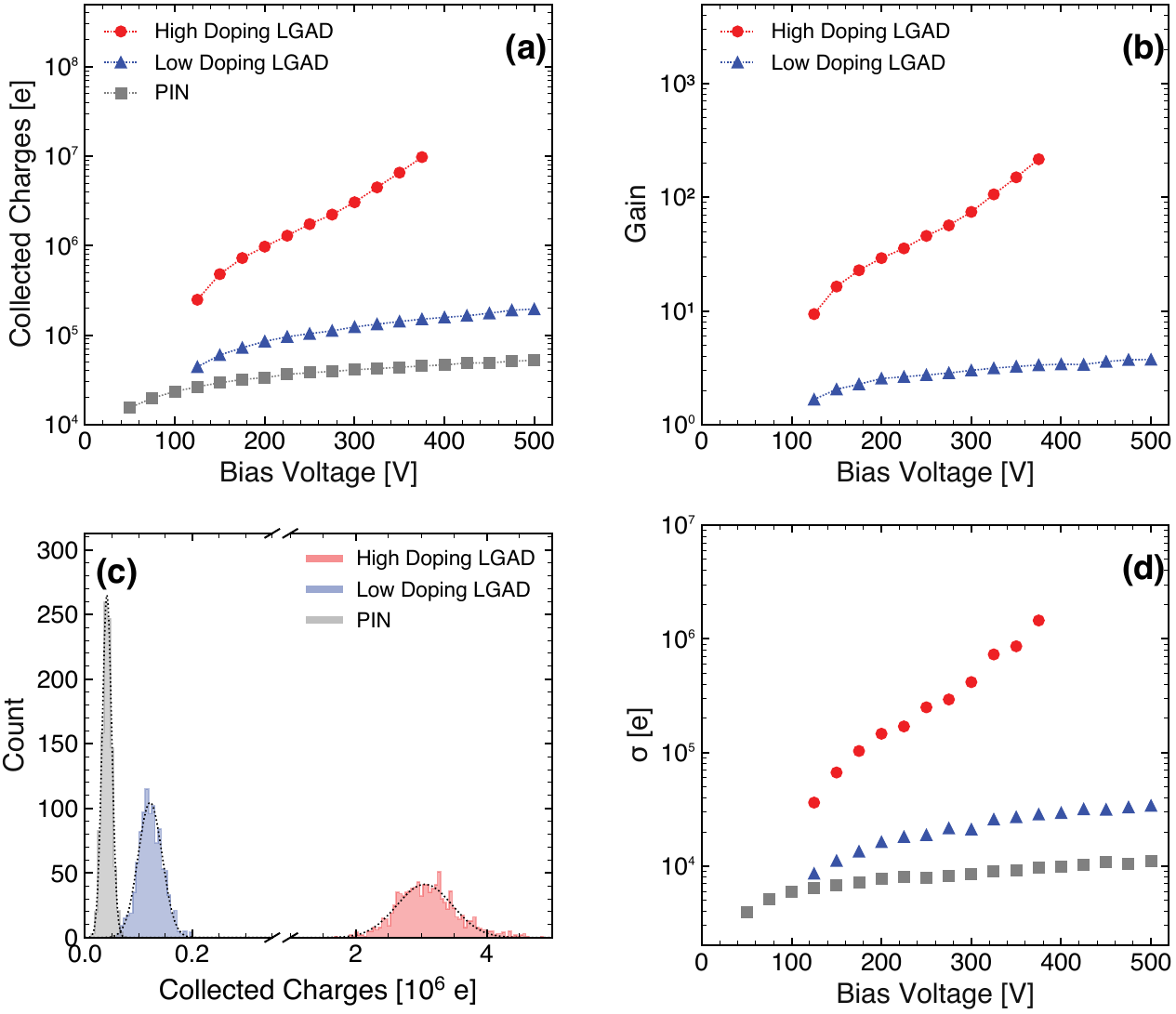}}
\caption{(a) Collected charge versus bias voltage for the high-doping LGAD, the low-doping LGAD, and the PIN diode measured at laser intensity I4. (b) Gain of the two LGADs, defined as the ratio of the LGAD collected charge to the PIN collected charge at the same bias voltage. (c) Charge spectra at 300~V with Gaussian fits (dotted lines); (d) Standard deviation of the collected charge versus bias voltage.}
\label{fig:collected_charges}
\end{figure}

\begin{figure}[!t]
\centerline{\includegraphics[width=1.0\columnwidth]{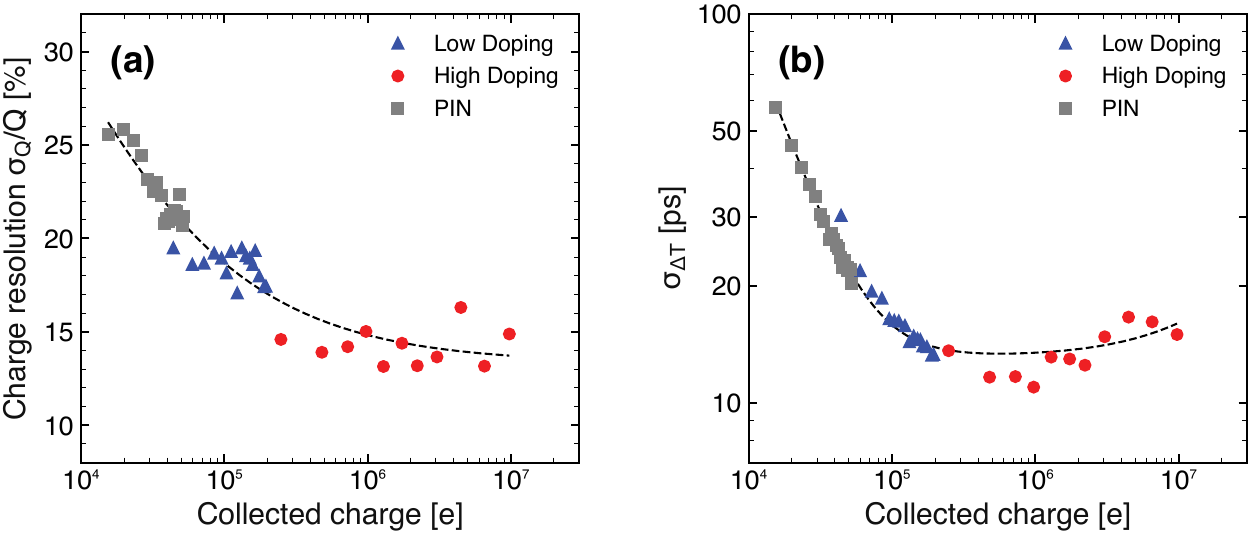}}
\caption{(a) Relative charge resolution $\sigma_Q/Q$ and (b) time resolution $\sigma_{\Delta T}$ as a function of the collected charge.}
\label{fig:resolution}
\end{figure}

Fig.~\ref{fig:collected_charges}(a) shows the collected charge as a function of bias for the three devices at the same laser intensity. The collected charge of the PIN diode increases slowly with the expansion of the depletion volume, from $1.6\times10^{4}$~e at 50~V to $5.2\times10^{4}$~e at 500~V, providing the unity-gain reference. This continued increase reflects that, as noted in Sec. II, the 75 μm drift layer is not fully depleted within the investigated bias range, so that the PIN diode and both LGADs remain partially depleted throughout the measurements. The lower-doped LGAD collects a moderately amplified charge, reaching $2.0\times10^{5}$~e at 500~V, whereas the higher-doped LGAD shows a nearly exponential increase from $2.5\times10^{5}$~e at 125~V to $9.8\times10^{6}$~e at 375~V. The corresponding gain, defined as the ratio of the LGAD charge to the PIN charge at the same bias, is shown in Fig.~\ref{fig:collected_charges}(b): the higher-doped device rises from 9.4 at 125~V to 216 at 375~V, while the lower-doped device remains below 3.7 up to 500~V. For what follows, the practical significance is that this pair of devices covers the range from unity gain to above 200 while the geometry, the epitaxial thickness, and the readout circuit remain unchanged.

The charge spectra of all three devices are Gaussian, as shown in Fig.~\ref{fig:collected_charges}(c) for 300 V, and the width of the collected charge distribution increases together with the bias and the gain (see Fig.~\ref{fig:collected_charges}(d)). The increasing width of the collected-charge distribution with gain is attributed to the stochastic nature of avalanche multiplication, which introduces excess noise and amplifies event-to-event charge fluctuations. Consequently, the Gaussian distribution becomes broader at higher gain \cite{Diff_Source_Test,Excess_Noise}.
The relative charge resolution $\sigma_Q/Q$ as a function of the collected charge is shown in Fig.~\ref{fig:resolution}(a). Using the collected charge as the horizontal axis places the three devices on a common scale and shows that $\sigma_Q/Q$ improves monotonically with increasing signal size: from about 25\% for the PIN diode at the lowest bias to about 14\% for the higher-doped LGAD at the largest signal. The improvement is steepest below $10^{5}$~e and largely saturates above $10^{6}$~e.

\section{Gain Dependence of the Time Resolution} \label{sec:timing}

\begin{figure}[!t]
\centerline{\includegraphics[width=1.0\columnwidth]{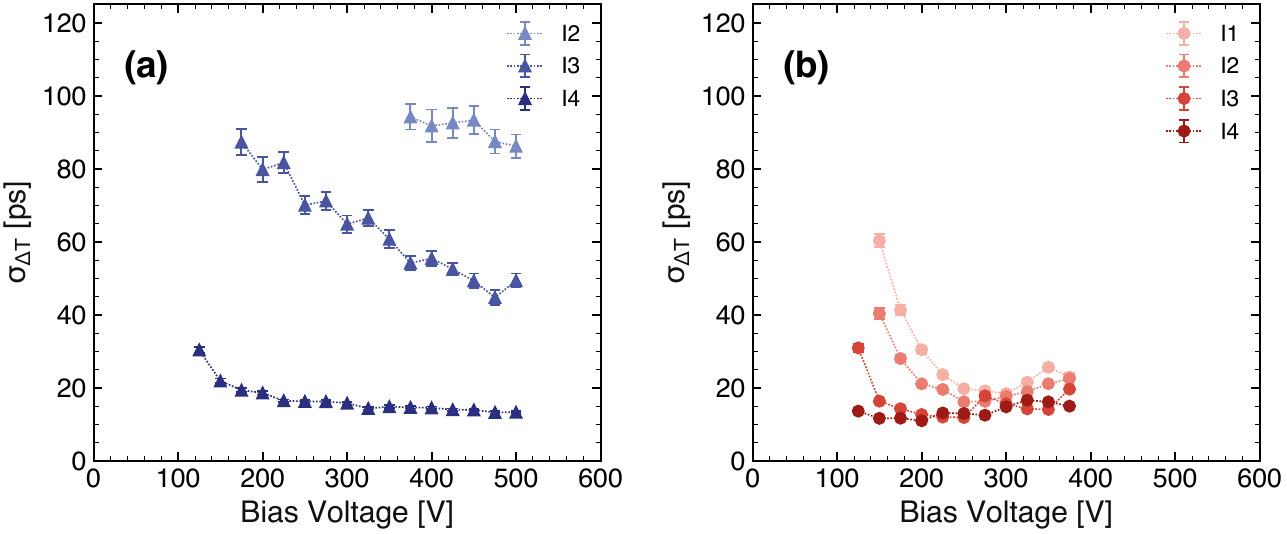}}
\caption{Time resolution $\sigma_{\Delta T}$ versus bias voltage at four laser intensities (I1-I4, increasing) for (a) the low-doping LGAD and (b) the high-doping LGAD.}
\label{fig:time_resolution}
\end{figure}

\begin{figure}[!t]
\centerline{\includegraphics[width=1.0\columnwidth]{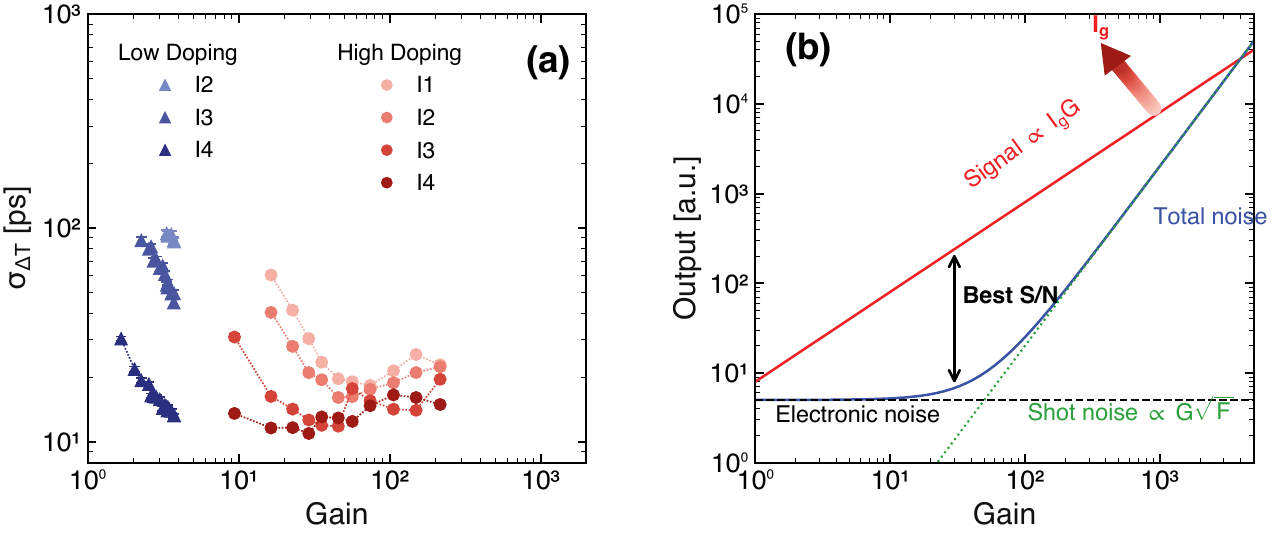}}
\caption{(a) Measured time resolution as a function of gain for all intensity series. The gain on the horizontal axis is determined from the I4 measurement and applied to all laser intensities. (b) Qualitative signal-to-noise picture, Ref.~\cite{Book_UFSD}: the signal grows in proportion to the gain, while the noise stays at the gain-independent electronics floor until the multiplied shot noise, which scales as $G\sqrt{F}$ with $F$ the excess noise factor, becomes comparable to it. The signal-to-noise ratio is therefore best at an intermediate gain, and the jitter $\sigma_n/(dV/dt)$ passes through a corresponding minimum. Increasing the injected charge $I_g$ raises the signal line and improves the ratio everywhere.}
\label{fig:sn}
\end{figure}

The time resolution $\sigma_{\Delta T}$ is defined as the standard deviation of the time difference between the LGAD signal and the laser trigger reference, both processed with a 50\% constant-fraction discriminator. The absolute values of $\sigma_{\Delta T}$ reported below are obtained under laser injection conditions that differ considerably from those of minimum ionizing particles, and therefore do not represent the actual MIP case. The quantity of interest here is the dependence of $\sigma_{\Delta T}$ on gain and injected charge. Fig.~\ref{fig:time_resolution} shows $\sigma_{\Delta T}$ as a function of bias at the four laser intensities, from which two trends can be identified. First, at a fixed illumination the higher-doped LGAD resolves better than the lower-doped device over the measured range: at intensity I2, for example, the lower-doped LGAD reaches only about 90~ps even at its maximum applied bias, whereas the higher-doped device remains at about 16 ps. Second, for the higher-doped device the resolution does not improve monotonically with bias: beyond a moderate bias the curve first flattens and then degrades. It should be noted that at intensity I1 the illumination is so weak that no signal can be observed in the lower-doped LGAD.

When the same data are plotted against gain, a common structure emerges. Owing to the low laser intensities, a reliable PIN-based calibration is only available at the highest intensity. Therefore, the gain of each LGAD at a given bias is defined from the I4 data set, and this gain value is applied uniformly to all laser intensities. As shown in Fig.~\ref{fig:sn}, each intensity series exhibits a minimum in $\sigma_{\Delta T}$ within a gain window of about 20 to 80 and deteriorates at higher gain. This behavior agrees qualitatively with the signal-to-noise picture given in \cite{Book_UFSD} for Si LGADs and illustrated in Fig.~\ref{fig:sn}(b). The jitter contribution to the time resolution is  $\sigma_{j} = \sigma_{n}/(dV/dt)$. The signal grows in proportion to the gain $G$, whereas the noise at low gain is dominated by the gain-independent readout electronics. The shot noise associated with the leakage current is itself multiplied and grows as $G\sqrt{F}$, where the excess noise factor $F$ reflects the stochastic nature of the avalanche process, and it eventually catches up with and surpasses the electronic term. The signal-to-noise ratio is therefore optimal at moderate gain, and the jitter passes through a corresponding minimum. Increasing the injected charge raises only the signal without changing the noise, shifting the entire curve downward, which is consistent with the observation in Fig.~\ref{fig:sn}(a). The measurements agree qualitatively with this description: the resolution improves steadily while electronic noise dominates, flattens where the two noise terms become comparable, and degrades once the multiplication shot noise takes over.

\section{Verification with a Beta Source} \label{sec:beta}

\begin{figure}[!t]
\centerline{\includegraphics[width=1.0\columnwidth]{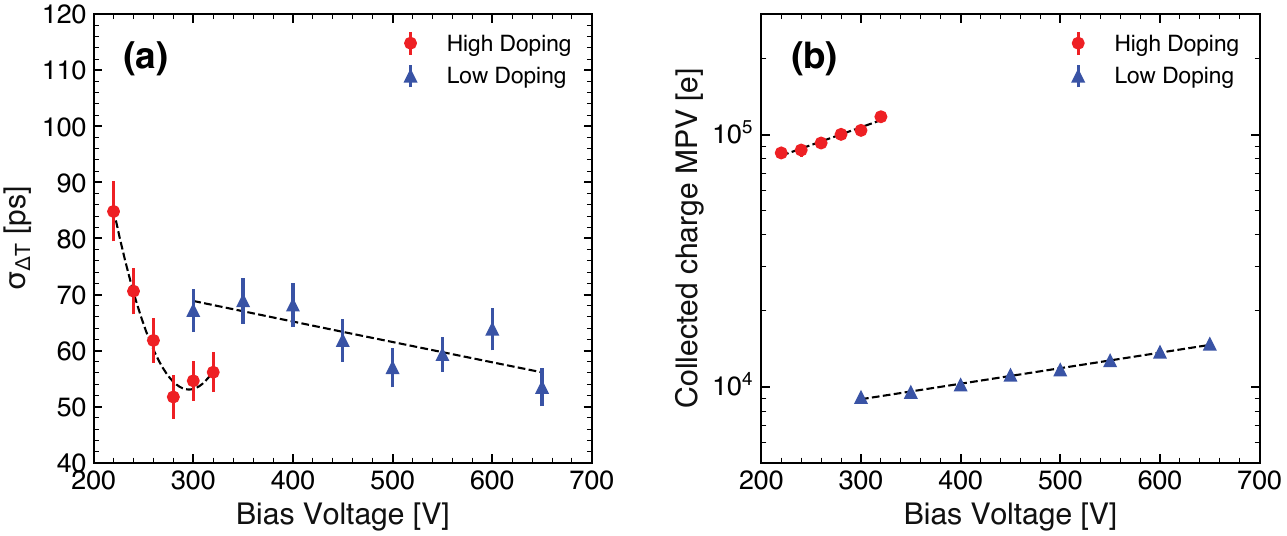}}
\caption{(a) Time resolution and (b) collected charge of the two 4H-SiC LGADs measured with a $^{90}$Sr beta source, against a Si LGAD reference held at 250~V. The reference contribution of $41.6 \pm 1.2$~ps has been subtracted in quadrature from (a). The high-doping device is shown up to 320~V, the bias beyond which a growing fraction of its pulses exceeds the oscilloscope range.}
\label{fig:beta}
\end{figure}

The charge injection in the UV-TCT measurements above differs considerably from that of a minimum ionizing particle and therefore does not represent the actual MIP conditions, so the two devices are also measured under a $^{90}$Sr beta source to characterize their detection capability for MIPs. Following \cite{Tao_TNS_Sr90}, a 50~${\mu}$m Si LGAD from Brookhaven National Laboratory (BNL) \cite{BNL_LGAD}, operated at 250 V, serves as a timing reference and is read out through the same amplifier and oscilloscope. The time difference between the two signals is processed with a constant-fraction discriminator, and the resolution of the reference detector, $41.6 \pm 1.2$~ps,  is subtracted in quadrature. The charge is obtained from a Landau fit to events falling within a $\pm 3\sigma$ coincidence window.

Fig.~~\ref{fig:beta} shows the measurement results. The higher-doped LGAD improves from $84.8$~ps at 220~V to $51.7$~ps at 280~V, whereas the lower-doped device requires more than 450 V to enter the same range and reaches $53.4$~ps only at 650~V. The collected charges at their respective optimal operating points differ by about a factor of eight.

The beta source data also reproduce, under particle excitation, the non-monotonic behavior observed in the laser measurements. The resolution of the higher-doped device passes through a minimum at 280 V and then rises to $56.1$~ps at 320~V, indicating that the timing performance begins to degrade before the highest usable bias is reached.

\section{Conclusion}

In this work, two 4H-SiC LGADs differing only in the gain layer doping concentration, together with a PIN diode taken from the same epitaxial wafer as the reference, are used to map the time resolution as a function of gain with a 375 nm UV-TCT system. A 25\% difference in doping produces a gain difference of two orders of magnitude between the two devices, so that this pair covers the range from unity gain to above 200 while the geometry and the readout circuit remain otherwise identical.

Within this range, the time resolution varies non-monotonically with gain at all the injected charge levels studied and passes through a minimum at a gain of about 20 to 80. This behavior arises because the jitter determined by the electronic noise decreases as the signal grows, while the multiplication shot noise increases with gain. Beyond this window, the timing performance is observed to degrade rather than improve, and the breakdown margin also shrinks, as indicated by the reverse I-V characteristics of the higher-doped device. The relative charge resolution improves with the signal size and reaches about 14\% for the higher-doped device.

Measurements with a $^{90}$Sr beta source provide complementary verification under particle excitation. The high-doping LGAD reaches a time resolution of 51.7~ps at 280~V and deteriorates at higher bias, while the low-doping device achieves 53.4~ps at 650~V. The non-monotonic timing response of the high-doping device is consistent with the UV-TCT results, supporting the existence of an optimal operating point. The gain window identified here is specific to the investigated devices and UV-TCT conditions. The beta-source result of 51.7~ps does not establish a fundamental timing limit for 4H-SiC LGADs; further optimization of drift-layer doping and thickness, electric-field distribution, capacitance, and readout electronics remains to be explored.



\bibliographystyle{ieeetr}
\bibliography{p4_sic_lgad}

\end{document}